\documentclass{aa}  

\usepackage{txfonts}

\usepackage{multirow}
\usepackage{array}
\usepackage{amssymb}
\usepackage{graphicx}
\usepackage{amsmath}
\usepackage{enumitem}
\usepackage{lscape}
\usepackage[T1]{fontenc} 
\usepackage{wrapfig}
\usepackage{float}
\usepackage{siunitx}
\usepackage{dcolumn}

\usepackage{booktabs}

\usepackage{color}
\usepackage{natbib,twoopt}
\usepackage[hyphenbreaks]{breakurl}
\usepackage[breaklinks]{hyperref}      
\bibpunct{(}{)}{;}{a}{}{,}             
\definecolor{cobalt}{rgb}{0.06, 0.2, 0.65}
\hypersetup{
  colorlinks,
  citecolor=cobalt,
  linkcolor=[rgb]{0.8, 0.2, 1.0},
  urlcolor=cobalt,
}
\makeatletter
  \newcommandtwoopt{\citeads}[3][][]{\href{http://adsabs.harvard.edu/abs/#3}%
    {\def\hyper@linkstart##1##2{}%
     \let\hyper@linkend\@empty\citealp[#1][#2]{#3}}}
  \newcommandtwoopt{\citepads}[3][][]{\href{http://adsabs.harvard.edu/abs/#3}%
    {\def\hyper@linkstart##1##2{}%
     \let\hyper@linkend\@empty\citep[#1][#2]{#3}}}
  \newcommandtwoopt{\citetads}[3][][]{\href{http://adsabs.harvard.edu/abs/#3}%
    {\def\hyper@linkstart##1##2{}%
     \let\hyper@linkend\@empty\citet[#1][#2]{#3}}}
  \newcommandtwoopt{\citeyearads}[3][][]%
    {\href{http://adsabs.harvard.edu/abs/#3}
    {\def\hyper@linkstart##1##2{}%
     \let\hyper@linkend\@empty\citeyear[#1][#2]{#3}}}
\makeatother

\begin{document} 

\title{Stellar mass growth in COSMOS-Web: a mass-complete main sequence to $z \sim 8$ and its consistency with GSMF evolution}

\author{M.~P.~Koprowski\inst{\ref{inst:tor}}
\and
K.~Lisiecki\inst{\ref{inst:tor},\ref{inst:ncbj}}
\and
P.~Sawant\inst{\ref{inst:tor}}
\and
J.~V.~Wijesekera\inst{\ref{inst:tor}}
        }

\institute{
Institute of Astronomy, Faculty of Physics, Astronomy and Informatics, Nicolaus Copernicus University, Grudzi\c{a}dzka 5, 87-100 Toru\'{n}, Poland, {\tt drelkopi@gmail.com}\label{inst:tor}
\and
National Centre for Nuclear Research, Pasteura 7, 02-093, Warsaw, Poland \label{inst:ncbj}
}

\date{}

\abstract
{The star-forming main sequence (MS) links the instantaneous star-formation rate of galaxies to their stellar-mass growth and therefore to the evolution of the galaxy stellar mass function (GSMF).}
{We determine a mass-complete MS in COSMOS-Web out to $z\simeq8$ and test whether its mass and redshift dependence is consistent with the observed evolution of the GSMF.}
{We stacked 251,462 star-forming galaxies in \textit{Herschel} and JCMT maps and corrected the far-infrared fluxes for blending using forward-modeled maps. We combined the resulting infrared luminosities with the unobscured ultraviolet emission to derive total star-formation rates and fitted a redshift-dependent MS. We independently measured the quiescent fraction and evolved the observed $z\simeq8$ COSMOS-Web GSMF forward within a continuity-equation framework including stellar-mass return, suppression of in-situ growth by the quiescent fraction, and mergers.}
{The MS is well described by a power law at low masses with a redshift-dependent turnover toward high masses. We measure a low-mass slope of $\gamma=1.215\pm0.028$, implying a mildly increasing specific star-formation rate with stellar mass. The GSMF evolved using this relation follows the observed progressive flattening of the low-mass end substantially better than the literature MS prescriptions considered here. Mergers change the low-mass number densities by $\lesssim0.2$ dex by $z\simeq0.65$ and have little effect on the slope. Small differences in MS normalization also accumulate strongly, producing large discrepancies in the evolved GSMFs     after several Gyr.}
{The observed GSMF evolution provides an independent integral test of the MS. Our results favor a mildly super-linear low-mass MS and show that both its slope and normalization must be measured accurately to reproduce the buildup of the galaxy stellar-mass distribution.}


\titlerunning{The COSMOS-Web main sequence and GSMF evolution}
\authorrunning{Koprowski et al.}

\maketitle

\section{Introduction}
\label{sec:intro}

Star-forming galaxies occupy a relatively tight relation between star-formation rate, $\mathrm{SFR}$, and stellar mass, $M_\ast$, commonly referred to as the star-forming main sequence (MS). This relation has been measured over a wide range of redshifts using rest-frame ultraviolet (UV), optical, infrared (IR), and radio star-formation indicators \citep[e.g.][]{Noeske_2007,Daddi_2007,Elbaz_2007,Whitaker_2012,Speagle_2014,Schreiber_2015,Koprowski_2024,Merida_2026}. Its existence implies that the growth of most star-forming galaxies is dominated by sustained star formation rather than short-lived starburst episodes \citep[e.g.][]{Rodighiero_2011}. The MS, therefore, provides a measure of the specific star-formation rate, $\mathrm{sSFR}\equiv\mathrm{SFR}/M_\ast$, and hence of the rate at which galaxies build their stellar mass through in-situ star formation.

The shape of the star-forming main sequence is particularly important for galaxy evolution, since its normalization sets how quickly galaxies grow in stellar mass, while its mass dependence determines how this growth varies with stellar mass. A nearly linear relation corresponds to an approximately mass-independent sSFR, whereas departure from linearity implies different fractional stellar-mass growth rate between star-forming galaxies. At high masses, the observed flattening of the MS implies a decreasing growth rate and therefore affects the buildup of the massive end of the galaxy stellar-mass function (GSMF). Integrated over cosmic time, even relatively small differences between main-sequence prescriptions can consequently produce substantial differences in the evolving stellar-mass distribution.

Measuring this relation, however, becomes increasingly difficult at high redshift. Rest-frame UV and optical data provide large samples and well-constrained stellar masses, but the inferred SFRs depend sensitively on dust-attenuation corrections. FIR and submillimeter observations more directly constrain obscured star formation, but individual detections are generally restricted to the most luminous systems because of sensitivity, confusion, and angular-resolution limitations. As a result, many high-redshift MS measurements rely on SED-derived SFRs, heterogeneous combinations of star-formation indicators, or IR-selected samples that preferentially contain massive and luminous galaxies \citep[e.g.][]{Speagle_2014,Schreiber_2015,Popesso_2023,Merida_2026}. Statistical approaches such as stacking can overcome the FIR detection limit \citep[e.g.][]{Tomczak_2016,Koprowski_2024}, but require careful treatment of source blending and a mass-complete parent population if the intrinsic MS shape is to be recovered.

James Webb Space Telescope (JWST) provides a major improvement by enabling deep, high-resolution rest-frame optical selection of galaxies far into the early Universe \citep{Gardner_2023}. COSMOS-Web combines JWST/NIRCam and MIRI imaging with the extensive multi-wavelength coverage of the COSMOS field and provides photometric redshifts, stellar masses, rest-frame colors, and other physical parameters for almost a million sources \citep{Casey_2023,Shuntov_2025}. This makes it possible to define large mass-complete samples of star-forming and quiescent galaxies over a substantially wider stellar-mass and redshift range than was accessible in previous studies. JWST does not, however, directly measure the cold-dust emission, making complementary FIR and submillimeter observations essential for a direct determination of star-forming galaxies' total SFRs.

The connection between the MS and stellar-mass assembly has previously been investigated by integrating observed MS relations through cosmic time. Main-sequence integration has been used to reconstruct average stellar-mass growth and star-formation histories \citep[e.g.][]{Noeske_2007b,Leitner_2012}. At the population level, \citet{Leja_2015} tested whether observed MS relations could reproduce the measured evolution of the GSMF. They showed that the shape of the relation, especially at low stellar masses, strongly affects the predicted stellar-mass function evolution. They also derived a main sequence relation that was directly constrained by the stellar-mass function itself and tested whether mergers or an additional population of quiescent galaxies could explain the apparent differences. Other continuity-equation approaches have instead started from observed SFR functions and assumed star-formation histories to predict evolving GSMF \citep{Lapi_2017}, or combined MS growth, transitions to quiescence, and mergers in empirical population models \citep{Steinhardt_2017}. A simpler test was also presented by \citet{Koprowski_2024}, demonstrating that differences between published MS prescriptions can become strongly amplified when integrated over time.

Our approach follows the same general continuity argument but differs from previous work in several important respects: the MS and quiescent fraction are measured independently from the same COSMOS-Web population, with the obscured star-formation component constrained directly using \textit{Herschel} and James Clerk Maxwell Telescope (JCMT) data to derive $L_{\rm IR}$; the observed GSMF is used only as an external consistency test rather than to constrain the MS; mergers are treated through both accreted stellar mass and donor destruction; and exactly the same population-evolution framework is applied to several literature MS prescriptions. The structure of the paper is the following. In Sect.~\ref{sec:data}, we define a mass-complete COSMOS-Web galaxy sample, adopt the corresponding stellar masses and redshifts, and introduce the FIR data and observed COSMOS-Web GSMFs used in the analysis. In Sect.~\ref{sec:meth}, we stack the \textit{Herschel} and JCMT maps and correct the recovered fluxes for blending using forward-modeled FIR maps. We then derive a mass-complete star-forming MS out to $z=8$ in Sect.~\ref{sec:ms}, using FIR+UV SFRs, and independently determine the mass- and redshift-dependent quiescent fraction from the same COSMOS-Web population in Sect.~\ref{sec:qg}. In Sect.~\ref{sec:consist}, the observed COSMOS-Web GSMF at $z\simeq8$ is evolved forward using the measured MS as the in-situ growth term, the independently measured $f_{\rm Q}(M_\ast,z)$ to suppress the population-averaged star-forming contribution, stellar-mass return, and a merger prescription that includes both accreted stellar mass and the corresponding destruction of donor galaxies; the complete continuity-equation formulation and numerical implementation are described in Appendix~\ref{sec:consist_pipe}. We discuss the resulting MS shape, its comparison with previous determinations, and the implications for stellar-mass growth in Sect.~\ref{sec:disc}, and summarize our conclusions in Sect.~\ref{sec:summ}. Throughout the paper, we adopt the \citet{Chabrier_2003} stellar IMF and a spatially flat Planck 2018 cosmology \citep{Planck_2020}, with $H_0=67.66\,{\rm km\,s^{-1}\,Mpc^{-1}}$ and $\Omega_{\rm m}=0.30966$.

\section{Data}
\label{sec:data}

\subsection{COSMOS-Web catalog}
\label{sec:data_cat}

We used version 1.1 of the COSMOS-Web master catalog presented by \citet{Shuntov_2025}. The catalog is based on the 255-h COSMOS-Web survey, which provides JWST/NIRCam imaging in the F115W, F150W, F277W, and F444W filters over $\sim$0.54\,$\mathrm{deg}^{2}$, together with MIRI/F770W imaging over $\sim$0.2\,$\mathrm{deg}^{2}$. Sources were detected from a combination of the four NIRCam images using a combined `hot and cold' procedure, with cold run used for bright/extended galaxies and hot run for faint/compact sources. The JWST observations were then combined with the extensive HST and ground-based imaging available in COSMOS, with the corresponding photometry measured with \texttt{SourceXtractor++} \citep{Bertin_2020,Kummel_2022} across 37 bands spanning approximately $0.3$--$8$\,\si{\micro\meter}. The catalog-level quantities used in this work, including photometric redshifts, stellar masses, rest-frame colors, and their associated uncertainties, were adopted directly from \citet{Shuntov_2025}.

Photometric redshifts were derived by fitting the multi-band total photometry with \texttt{LePhare} \citep{Arnouts_1999,Ilbert_2006}, based on the \citet{Bruzual_2003} stellar-population models, where exponentially declining and delayed star-formation histories, two stellar metallicities, nebular emission lines, and several dust-attenuation prescriptions were assumed. In this work we adopted the median of the redshift probability distribution, with the corresponding 16th and 84th percentiles defining the redshift errors. As explained in \citet{Shuntov_2025}, comparison with high-quality spectroscopic redshifts gives $\sigma_{\mathrm{MAD}}=0.012$ for sources with $m_{\mathrm{F444W}}<28$, increasing to approximately $0.03$ for the faintest sources.

Stellar masses were initially derived from the photometric redshifts run using the same \texttt{LePhare} template library. The physical parameters were then recalibrated, with the redshift fixed to the adopted median estimate. Throughout this work, we used the stellar masses from this fixed-redshift fit, together with their catalog-provided 16th and 84th percentile uncertainties. These mass errors, therefore, describe the uncertainty of the SED fit at fixed redshift, while the photometric-redshift errors were propagated separately in our analysis. The stellar-mass completeness limits applied to the final sample are described in Sect.~\ref{sec:data_selec}.

\subsection{Ancillary FIR data}
\label{sec:data_FIR}

To measure the dust-obscured star formation of the COSMOS-Web galaxies, we used far-infrared imaging of the COSMOS field obtained with the \textit{Herschel Space Observatory} \citep{Pilbratt_2010}. The adopted data comprise the 100 and 160\,\si{\micro\meter} PACS maps from the PACS Evolutionary Probe survey \citep[PEP;][]{Poglitsch_2010,Lutz_2011} and the 250, 350, and 500\,\si{\micro\meter} SPIRE maps from the \textit{Herschel} Multi-tiered Extragalactic Survey \citep[HerMES;][]{Griffin_2010,Oliver_2012}. The beam full widths at half maximum (FWHM) are 7.39, 11.29, 18.2, 24.9, and 36.3\,arcsec at 100, 160, 250, 350, and 500\,\si{\micro\meter}, respectively, while the corresponding nominal $5\sigma$ sensitivities are 7.7, 14.7, 24.0, 27.5, and 30.5\,mJy.

We complemented the \textit{Herschel} imaging with the 850\,\si{\micro\meter} map obtained with SCUBA-2 on JCMT \citep{Holland_2013} as part of the SCUBA-2 Cosmology Legacy Survey \citep[S2CLS;][]{Geach_2017}. The COSMOS S2CLS imaging has a beam FWHM of approximately 14.8\,arcsec and a typical $1\sigma$ depth of approximately 1.6\,mJy\,beam$^{-1}$. The resulting coverage from 100 to 850\,\si{\micro\meter} provides leverage on both the peak and the long-wavelength side of the dust SED across the redshift range considered here, allowing the stacked flux densities to constrain the total infrared luminosity more robustly than measurements based on a single FIR band. The stacking and flux-extraction procedures applied to these maps are described in Sect.~\ref{sec:meth}.

\subsection{Sample selection}
\label{sec:data_selec}

We began by applying the following catalog-quality and galaxy-selection criteria recommended for COSMOS-Web \citep{Shuntov_2025}. From the 784,016 items in the COSMOS-Web v1.1 master catalog, we removed sources that were either affected by the JWST/NIRCam stellar masks, hot pixels, strongly blended in the F444W image, or classified by the SED fitting as stars or quasars. In addition, we excluded detections that had inconsistent ground- and space-based photometry, unrealistically small measured sizes, anomalous aperture-flux ratios or were confined to a single NIRCam band. We also required valid positive estimates of the photometric redshift and its probability distribution. These initial criteria leave 648,813 objects.

Following \citet{Shuntov_2025_GSMF}, we additionally removed sources with poorly constrained photometric redshifts by requiring more than 25\% of the redshift probability to lie within $z_{\rm phot}\pm\Delta z$, where $\Delta z$ is the width of the corresponding redshift bin. Approximating the redshift probability distribution as a Gaussian with $\sigma_z=(z_{84}-z_{16})/2$, this criterion was implemented as 

\begin{equation}
{\rm erf}\left(\frac{\Delta z_{\rm bin}}{\sqrt{2}\,\sigma_z}\right)>0.25,
\label{eq:z_probability_selection}
\end{equation}

\noindent and left 487,654 sources between the adopted redshift range of $0.5\leq z<8.0$. In addition, the sample was limited to objects with the best-fitting galaxy template producing lower $\chi^2$ than the best-fitting stellar template. At $z>3.5$, we also removed possible compact red AGN and little-red-dot contaminants following the selection adopted by \citet{Akins_2025}. These comprise compact sources for which either the AGN template provides a better fit than the galaxy template or the observed color satisfies $m_{\rm F277W}-m_{\rm F444W}>1.5$. After these additional cuts, 479,388 sources remain.

For the stacking purposes, the sample was divided into the redshift intervals $0.5\leq z<1.0$, $1.0\leq z<1.5$, $1.5\leq z<2.0$, $2.0\leq z<3.0$, $3.0\leq z<5.0$, and $5.0\leq z<8.0$. The stellar-mass bins have a width of 0.75 dex, with edges at $\log(M_\ast/{\rm M_\odot})=7.5$, 8.25, 9.0, 9.75, 10.5, and 11.25. We adopted lower stellar-mass completeness limits of $\log(M_\ast/{\rm M_\odot})_{\rm lim}=7.5$, 7.5, 8.25, 8.25, 9.0, and 9.0 in the respective redshift intervals, as determined by \citet{Shuntov_2025}, which produces a mass-complete parent sample of 268,276 sources.

Star-forming and quiescent galaxies were separated using the rest-frame $NUV-r-J$ colors derived from the catalog SED fits. Following \citet{Ilbert_2013}, an object was classified as quiescent when:

\begin{equation}
\begin{aligned}
(NUV-r) &> 3.0\times (r-J)+1.0, \\
(NUV-r) &> 3.1.
\end{aligned}
\label{eq:nuvrj_selection}
\end{equation}

\noindent This classification identifies 16,559 quiescent and 251,717 star-forming galaxies. Quiescent sources were excluded from the stacking sample used to measure the star-forming main sequence. The evolution of the quiescent fraction was examined separately in Sect.~\ref{sec:qg}.

Finally, in order to remove starbursts, we cross-matched the star-forming sample with the available A3COSMOS ALMA catalog \citep{Liu_2019} using a matching radius of 1\,arcsec (299 counterparts found). Following \citet{Elbaz_2018}, 255 of them with ${\rm SFR_{ALMA}}>3 \times {\rm SFR_{MS}}$ were classified as starbursts, where the main sequence functional form of \citet{Koprowski_2024} was adopted. These objects were removed from the main-sequence stacking sample, leaving 251,462 galaxies. Because A3COSMOS is based on heterogeneous archival ALMA observations from individual PI-led programs and does not define a luminosity-complete sample, some starbursts may remain unidentified, the impact of which is discussed in Sect.~\ref{sec:disc_mass_grow}. The cumulative sample sizes and final population breakdown are summarized in Table~\ref{tab:sample_selection}.

\begin{table}
\begin{footnotesize}
\begin{center}
\caption{Numbers of COSMOS-Web sources retained during the sample selection. The first five rows give cumulative counts, while the remaining rows describe subsets of the mass-complete parent sample and the final star-forming stacking sample.}
\label{tab:sample_selection}
\setlength{\tabcolsep}{2.5mm}
\begin{tabular}{lr}
\hline\hline
Selection stage or population & Number \\
\hline
COSMOS-Web v1.1 master catalog              & 784,016 \\
Initial quality and galaxy selection          & 648,813 \\
Redshift range and redshift-quality selection & 487,654 \\
Stellar and compact-red-source rejection      & 479,388 \\
Mass-complete parent sample                   & 268,276 \\
\hline
Quiescent galaxies                            &  16,559 \\
Star-forming galaxies                         & 251,717 \\
ALMA counterparts                             & \phantom{000,}299 \\
ALMA-identified starbursts removed            & \phantom{000,}255 \\
Final main-sequence stacking sample           & 251,462 \\
\hline
\end{tabular}
\end{center}
\end{footnotesize}
\end{table}

\subsection{External GSMF comparison data}
\label{sec:data_gsmf}

For comparison with the galaxy stellar-mass functions evolved from our main-sequence measurements, we adopted the total GSMFs derived from the COSMOS-Web catalog by \citet{Shuntov_2025_GSMF}. These measurements were obtained from the same photometric catalog and the same stellar-mass estimates used in our analysis, thereby minimizing systematic differences associated with source selection, photometric redshifts, and SED modeling. The GSMFs were determined between $z=0.5$ and $z=12$, although only the measurements relevant to the redshift range considered in this work were used.

\citet{Shuntov_2025_GSMF} fitted the total GSMFs with double-Schechter functions in the lower-redshift intervals, where both components could be constrained, and with single-Schechter functions at $z>3.5$. We adopted their best-fitting characteristic masses, normalizations, and low-mass slopes, together with the corresponding asymmetric uncertainties. The analytical Schechter function in the highest-redshift interval was used to initialize the GSMF evolution, while the remaining redshift-dependent functions and their 16th--84th percentile ranges provided the observational reference against which the evolved GSMFs were compared in Sect.~\ref{sec:disc_mass_grow}.

\section{Stacking}
\label{sec:meth}

\subsection{Stacking procedure}
\label{sec:meth_stack}

We measured the average FIR emission of the selected star-forming galaxies by stacking in the PACS, SPIRE, and SCUBA-2 maps in the redshift and stellar-mass bins defined in Sect.~\ref{sec:data_selec}. For each galaxy, a map cutout centered on its COSMOS-Web position was extracted. The cutouts were combined using an inverse-variance-weighted mean, with the corresponding noise maps providing the pixel weights. The representative redshift and stellar mass in each bin were taken to be the mean values of the galaxies contributing to that stack.

We extracted the stacked flux densities from azimuthally averaged radial light profiles rather than from a single central pixel or a fixed aperture. Following \citet{Koprowski_2024}, the profiles were fitted with beam models corresponding to each instrument: single-Gaussian models for the PACS bands, double-Gaussian models for the SPIRE bands, and an empirical SCUBA-2 profile at 850\,\si{\micro\meter} \citep{Geach_2017}. The broader component of the SPIRE model was added in to account for the extended emission surrounding the compact central source. The same profile-extraction and fitting procedure was later applied to the simulated stacks used to estimate the blending correction (Sect.~\ref{sec:meth_deblend}).

This procedure produced an initial set of stacked flux densities at 100\,\si{\micro\meter}, 160\,\si{\micro\meter}, 250\,\si{\micro\meter}, 350\,\si{\micro\meter}, 500\,\si{\micro\meter}, and 850\,\si{\micro\meter} for each populated redshift and stellar-mass bin. The corresponding uncorrected $L_{\rm IR}$--$M_\ast$ relation (with the procedure for derivation of $L_{\rm IR}$ values described in Sect.~\ref{sec:ms_points}) showed an apparent flattening at the low-mass end, where blending is expected to have the largest effect. To determine whether this feature was caused by unresolved neighboring sources and to correct for it, we performed the simulated-map blending analysis described in Sect.~\ref{sec:meth_deblend}.

\subsection{Blending correction}
\label{sec:meth_deblend}

Stacking in confusion-limited FIR and submillimeter maps does not recover only the mean emission of the target galaxies. At low stellar masses, the GSMF implies a large number of galaxies on the sky, so several sources can fall within a single beam even without being physically related. Galaxy clustering adds a further contribution from nearby associated sources. Both effects increase the recovered stacked flux, with the strongest relative impact on the faintest, low-mass populations. If left uncorrected, this can bias their mean luminosities and change the measured slope of the main sequence. We therefore calibrated the combined effect using simulated FIR maps that preserved the observed positions and number density of the COSMOS-Web sources.

We initially used the redshift-dependent $L_{\rm IR}$--$M_\ast$ relation from the uncorrected stacked measurements (and later repeated the procedure using the corrected function; see Sect.~\ref{sec:ms_points} for $L_{\rm IR}$ derivation procedure), to assign an input FIR flux density to each star-forming galaxy according to its stellar mass and redshift (assuming the $T_{\rm d}-z$ relation derived in \citealt{Koprowski_2024}). For quiescent galaxies, the input FIR flux density was set to zero, while for the ALMA-matched sources, measured dust luminosities were used, assuming $T_{\rm dust}=30$\,K. Point sources were then injected into the simulated FIR maps at the observed sky coordinates of the COSMOS-Web galaxies. Although the ALMA-identified starbursts were excluded from the stacking sample, they remained present in the simulated maps because their emission can contribute to the flux measured around neighboring galaxies.

The simulated sources were then convolved with the corresponding instrumental beams. To remove the large-scale background, we smoothed each simulated map with a $\sigma=300\,\mathrm{arcsec}$ Gaussian kernel and subtracted it from the original map. We then repeated the complete stacking and flux-extraction procedure using the same source selection, redshift and stellar-mass bins, ALMA-starburst exclusion, and radial-profile fitting adopted for the real maps, explained in detail in Sect.~\ref{sec:meth_stack}. This ensured that any difference between the assigned input flux density, $S_{\rm input}$, and the recovered simulated values, $S_{\rm output}$, resulted from the map resolution, source clustering, and the stacking and extraction procedure.

We modeled the recovered simulated flux density as the sum of the input flux density and a non-negative, band-dependent blending contribution, $c_{\rm blend}$:

\begin{equation}
S_{\rm corr}=S_{\rm obs}-c_{\rm blend}.
\label{eq:blending_correction}
\end{equation}

\noindent Non-zero corrections were fitted for the 350\,\si{\micro\meter}, 500\,\si{\micro\meter}, and 850\,\si{\micro\meter} bands. Their values and corresponding uncertainties are listed in Table~\ref{tab:blending_offsets}. The offsets at 100\,\si{\micro\meter}, 160\,\si{\micro\meter}, and 250\,\si{\micro\meter} were found to be negligible and were therefore fixed to zero.

\begin{table}
\begin{footnotesize}
\begin{center}
\caption{Additive blending corrections derived from the simulated FIR maps in Sect.~\ref{sec:meth_deblend}.}
\label{tab:blending_offsets}
\setlength{\tabcolsep}{3.5mm}
\begin{tabular}{ccc}
\hline\hline
Wavelength &
$c_{\rm blend}$ &
$\sigma(c_{\rm blend})$ \\
 &
(mJy\,beam$^{-1}$) &
(mJy\,beam$^{-1}$) \\
\hline
350\,\si{\micro\meter} & 0.10869 & 0.02692 \\
500\,\si{\micro\meter} & 0.16803 & 0.02286 \\
850\,\si{\micro\meter} & 0.02353 & 0.00660 \\
\hline
\end{tabular}
\end{center}
\end{footnotesize}
\end{table}

\begin{figure*}
\centering
\includegraphics[width=\textwidth]{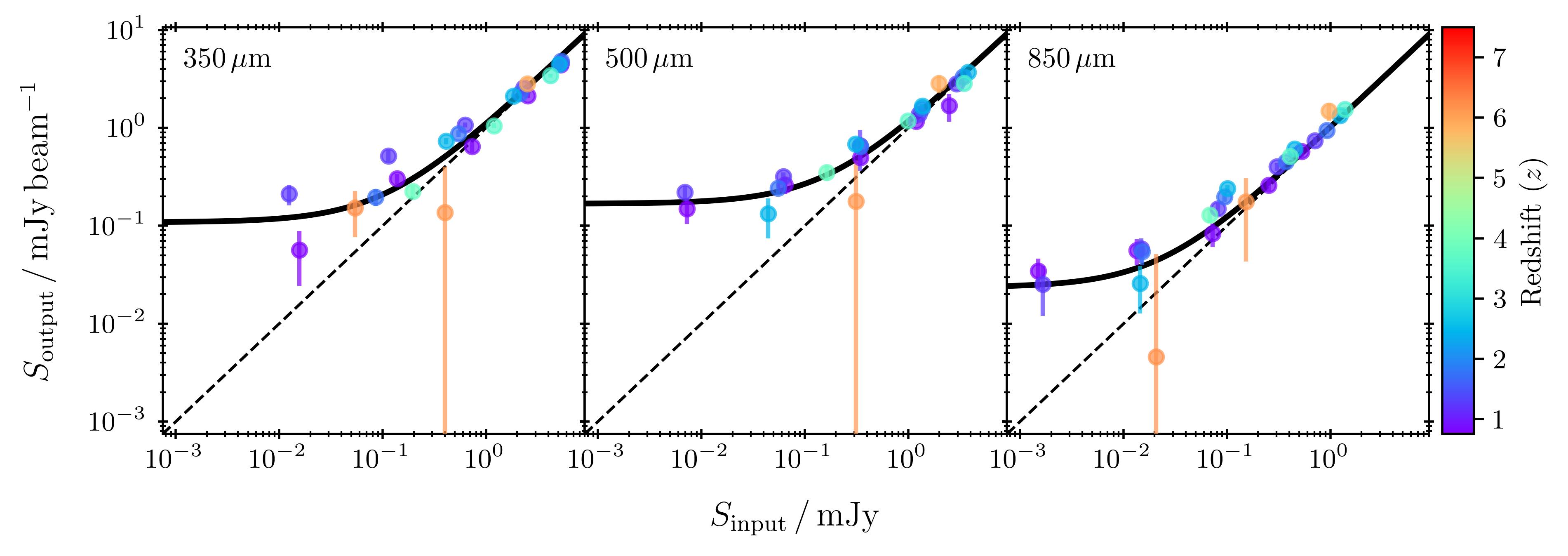}
\caption{Comparison between the input flux densities assigned to the simulated maps and the output flux densities recovered after applying the complete stacking and extraction procedure at 350, 500, and 850\,${\rm \mu m}$ (Sect.~\ref{sec:meth_deblend}). The points are color-coded by redshift. The dashed lines show the one-to-one relation, while the solid black lines show the fitted additive formulae ($S_{\rm output}=S_{\rm input}+c_{\rm blend}$). The separation between the fitted and one-to-one relations represents the band-dependent contribution from source blending.}
\label{fig:blending_correction}
\end{figure*}

Fig.~\ref{fig:blending_correction} compares the assigned and recovered flux densities in the three bands for which a blending correction was applied. The fitted relations describe the excess recovered flux as an additive contribution rather than a multiplicative scaling. This treatment reflects the fact that unresolved emission from neighboring sources adds flux within the instrumental beam and has the largest effect on the faintest stacked populations. The fitted offsets, $c_{\rm blend}$, were then subtracted from the corresponding observed stacked flux densities, with the resulting corrected numbers used for the final FIR SED fitting and the derivation of the intrinsic $L_{\rm IR}$--$M_\ast$ and SFR--$M_\ast$ relations, as described in Sect.~\ref{sec:ms}.

\subsection{Uncertainty estimation}
\label{sec:meth_errs}

The uncertainty of the stacked flux densities included the instrumental map noise, the uncertainty of the radial-profile fit, the redshift and $M_\ast$ errors, and, where applicable, the uncertainty of the blending correction. As explained in Sect.~\ref{sec:meth_stack}, the maps were stacked using an inverse-variance-weighted mean. At each pixel, the stacked signal and its uncertainty were calculated as

\begin{equation}
S_{\rm stack}
=
\frac{\sum_i w_i S_i}{\sum_i w_i},
\qquad
w_i=\frac{1}{\sigma_i^2},
\qquad
\sigma_{\rm stack}
=
\left(\sum_i w_i\right)^{-1/2},
\label{eq:stack_uncertainty}
\end{equation}

\noindent where $S_i$ and $\sigma_i$ are the signal and instrumental uncertainty, respectively, in the cutout of the $i$th galaxy.

Flux densities were measured by fitting the azimuthally averaged stacked profiles with the band-dependent beam models. The uncertainty on the extracted FIR flux densities was calculated as

\begin{equation}
\sigma_{\rm ext}
=
\left(
\sigma_{\rm central}^{2}
+
\sigma_{\rm fit}^{2}
\right)^{1/2},
\label{eq:extraction_uncertainty}
\end{equation}

\noindent where $\sigma_{\rm central}$ is the central-pixel value of the stacked uncertainty image and $\sigma_{\rm fit}$ is the error on the profile fit, derived from the corresponding covariance matrix.

We estimated the contribution from photometric-redshift and stellar-mass uncertainties using 1000 Monte Carlo realizations of the catalog. In each realization, the redshift and stellar mass of every galaxy were perturbed using Gaussian distributions, with the lower and upper widths determined from the catalog-provided 16th and 84th percentiles. Galaxies were then reassigned to the redshift and stellar-mass bins, and inverse-variance-weighted means of their fixed raw-map central-pixel flux densities were recalculated. The population standard deviation of the resulting flux-density distribution, $\sigma_{\rm MC}$, was combined with the profile-extraction uncertainty according to

\begin{equation}
\sigma_{\rm obs}
=
\left(
\sigma_{\rm ext}^{2}
+
\sigma_{\rm MC}^{2}
\right)^{1/2}.
\label{eq:observed_flux_uncertainty}
\end{equation}

\noindent For the 350\,\si{\micro\meter}, 500\,\si{\micro\meter}, and 850\,\si{\micro\meter} bands, the uncertainty of the additive blending correction was propagated into the final de-blended flux uncertainty:

\begin{equation}
S_{\rm corr}
=
S_{\rm obs}
-
c_{\rm blend},
\qquad
\sigma_{\rm corr}
=
\left[
\sigma_{\rm obs}^{2}
+
\sigma^{2}(c_{\rm blend})
\right]^{1/2}.
\label{eq:corrected_flux_uncertainty}
\end{equation}

\noindent The final de-blended FIR flux densities and their uncertainties for all redshift and stellar-mass bins used in the analysis are available in machine-readable form in the associated online material.

\section{The mass-complete star-forming main sequence out to z=8}
\label{sec:ms}

\subsection{Binned SFR--stellar-mass measurements}
\label{sec:ms_points}

We converted the final de-blended FIR flux densities into total infrared luminosities by fitting the stacked photometry in each redshift and stellar-mass bin with the modified-blackbody plus mid-infrared power-law model of \citet{Casey_2012}. Following \citet{Drew_2022}, the dust-emissivity index and mid-infrared power-law slope were fixed to $\beta=1.96$ and $\alpha=2.3$, respectively. The model was evaluated at the mean redshift of the galaxies contributing to each stack. Only bands with positive measured flux densities were used to determine the SED normalization and $L_{\rm IR}$.

The dust temperature was allowed to vary in bins with at least 3 ${\rm SNR}>1$ FIR bands. In addition, in order to trace both sides of the dust-emission peak, at $z<3$ detections were required at either 100\,\si{\micro\meter} or 160\,\si{\micro\meter} and at either 500\,\si{\micro\meter} or 850\,\si{\micro\meter}, while at $z>3$ the longest-wavelength eligible measurement had to be at 850\,\si{\micro\meter}. When these criteria were not satisfied simultaneously, or when the free-temperature fit failed, the temperature was fixed to the value in the nearest stellar-mass bin within the same redshift interval. The total infrared luminosity was calculated by integrating the fitted rest-frame SED between 8 and 1000\,\si{\micro\meter}. To propagate the photometric uncertainties, we generated Monte Carlo realizations in which each FIR flux density was drawn from a Gaussian distribution centered on its measured value with a standard deviation equal to its $1\sigma$ uncertainty. After refitting the SED, the uncertainty on $L_{\rm IR}$ was set to the $1\sigma$ dispersion of the individual MC values. The complete set of de-blended FIR SED fits for the individual redshift and stellar-mass bins is provided in the associated online material.

\begin{figure}
\centering
\includegraphics[width=\columnwidth]{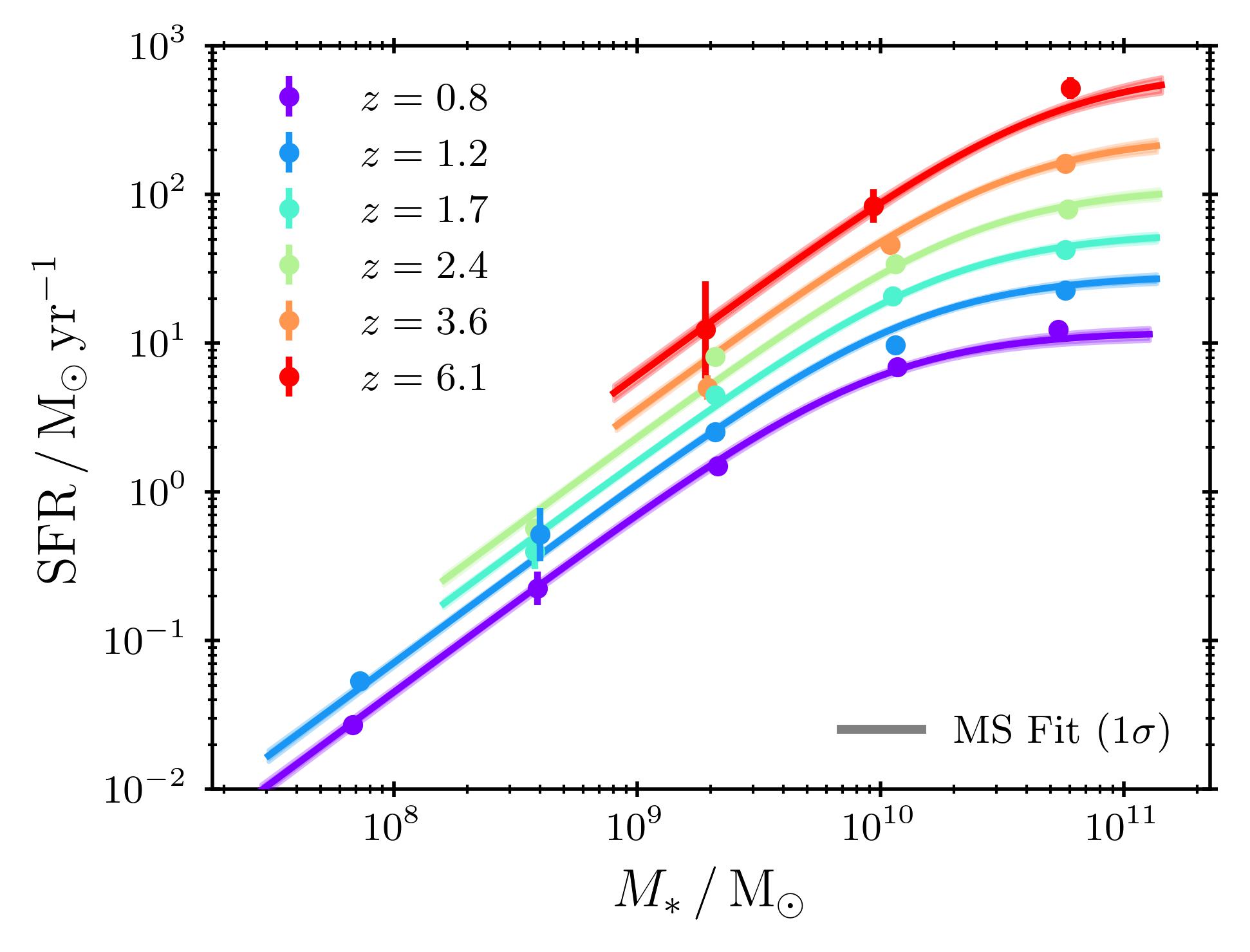}
\caption{Binned total SFR measurements as a function of stellar mass. The color-coded points and error bars show the measurements and their $1\sigma$ uncertainties in the six redshift bins (Sect.~\ref{sec:ms_points}) The solid curves show the best-fitting main-sequence relations, and the shaded regions indicate their $1\sigma$ uncertainties (Sect.~\ref{sec:ms_func}).}
\label{fig:ms_fit}
\end{figure}

To account for the unobscured star formation, we calculated the mean $L_{\rm NUV}$ in each redshift and stellar-mass bin from the individual catalog values. The corresponding uncertainty was determined from 1000 bootstrap resamples of the galaxies in each bin. The total SFR was then calculated following \citet{Kennicutt_2012}:

\begin{equation}
\mathrm{SFR} = 2.6\times10^{-10} \left(L_{\rm NUV} +
0.27L_{\rm IR}\right),
\label{eq:sfr_conversion}
\end{equation}

\noindent with luminosities expressed in ${\rm L_\odot}$ and the resulting SFR in ${\rm M_\odot}\,\mathrm{yr}^{-1}$. For bins in which the FIR SED fitting did not provide a valid $L_{\rm IR}$ measurement (not a single positive stacked FIR flux density), the obscured component and its uncertainty were set to zero, and the resulting SFR was therefore determined from $L_{\rm NUV}$ alone. The final total SFR measurements and their uncertainties are shown in Fig.~\ref{fig:ms_fit} as colored points with error bars. The binned $L_{\rm IR}$, $L_{\rm NUV}$, and SFR measurements, together with their uncertainties, are also available in machine-readable form in the associated online material.

\subsection{Main-sequence functional form}
\label{sec:ms_func}

Following \citet{Lee_2015}, we fitted the binned SFR--stellar-mass measurements with a smoothly broken relation that approaches a power law at low stellar masses and flattens above a redshift-dependent characteristic mass. Defining $x\equiv\log_{10}(M_\ast/{\rm M_\odot})$ and $\psi\equiv\log_{10}(\mathrm{SFR}/{\rm M_\odot}\,\mathrm{yr}^{-1})$, the adopted functional form was

\begin{equation}
\psi(x,z)
=
s_0(z)
-
\log_{10}
\left[
1+
10^{-\gamma\left[x-x_0(z)\right]}
\right],
\label{eq:ms_function}
\end{equation}

\noindent where $s_0(z)$ is the asymptotic high-mass logarithmic SFR, $x_0(z)\equiv\log_{10}(M_0/{\rm M_\odot})$ is the logarithmic turnover stellar mass, and $\gamma$ is the low-mass logarithmic slope. At $x\ll x_0$, the relation approaches $\psi\simeq s_0+\gamma(x-x_0)$, while at $x\gg x_0$ $\psi\simeq s_0$.

We described the redshift evolution of the normalization and turnover mass as linear functions of $\log_{10}(z)$,

\begin{equation}
\begin{aligned}
s_0(z) &= a_{s_0}\log_{10}(z)+b_{s_0},\\
x_0(z) &= a_{M_0}\log_{10}(z)+b_{M_0},
\end{aligned}
\label{eq:ms_redshift_evolution}
\end{equation}

\noindent while $\gamma$ was taken to be independent of redshift. The five parameters were fitted simultaneously to all valid mass-complete SFR measurements using a Gaussian likelihood and the affine-invariant ensemble sampler implemented in \texttt{emcee} \citep{Foreman_2013}. We adopted flat, bounded priors on all fitted parameters and initialized the walkers around the minimum-$\chi^2$ solution. The posterior samples were used to derive the best-fitting parameter values and their uncertainties given in Table~\ref{tab:ms_parameters}, and to calculate the $1\sigma$ uncertainty of the relation shown in Fig.~\ref{fig:ms_fit}. The fitted redshift evolution of $s_0$, $x_0$, and the best-fit value of $\gamma$, together with the corner plot, is provided in the associated online material. The posterior samples are also available in machine-readable form through the project GitHub repository.

\begin{table}
\begin{footnotesize}
\begin{center}
\caption{Best-fitting parameters of the redshift-dependent star-forming main-sequence model defined by Eqs.~\ref{eq:ms_function} and \ref{eq:ms_redshift_evolution}. The quoted errors are the $1\sigma$ posterior uncertainties.}
\label{tab:ms_parameters}
\setlength{\tabcolsep}{4.0mm}
\begin{tabular}{lc}
\hline\hline
Parameter & Best-fitting value \\
\hline
$a_{s_0}$ & $\phantom{0}2.002\pm0.112$ \\
$b_{s_0}$ & $\phantom{0}1.270\pm0.039$ \\
$a_{M_0}$ & $\phantom{0}0.789\pm0.131$ \\
$b_{M_0}$ & $10.072\pm0.064$ \\
$\gamma$  & $\phantom{0}1.215\pm0.028$ \\
\hline
\end{tabular}
\end{center}
\end{footnotesize}
\end{table}

\section{Quiescent galaxy fraction}
\label{sec:qg}

\subsection{Derivation of quiescent fractions}
\label{sec:qg_points}

\begin{figure*}
\centering
\includegraphics[width=\textwidth]{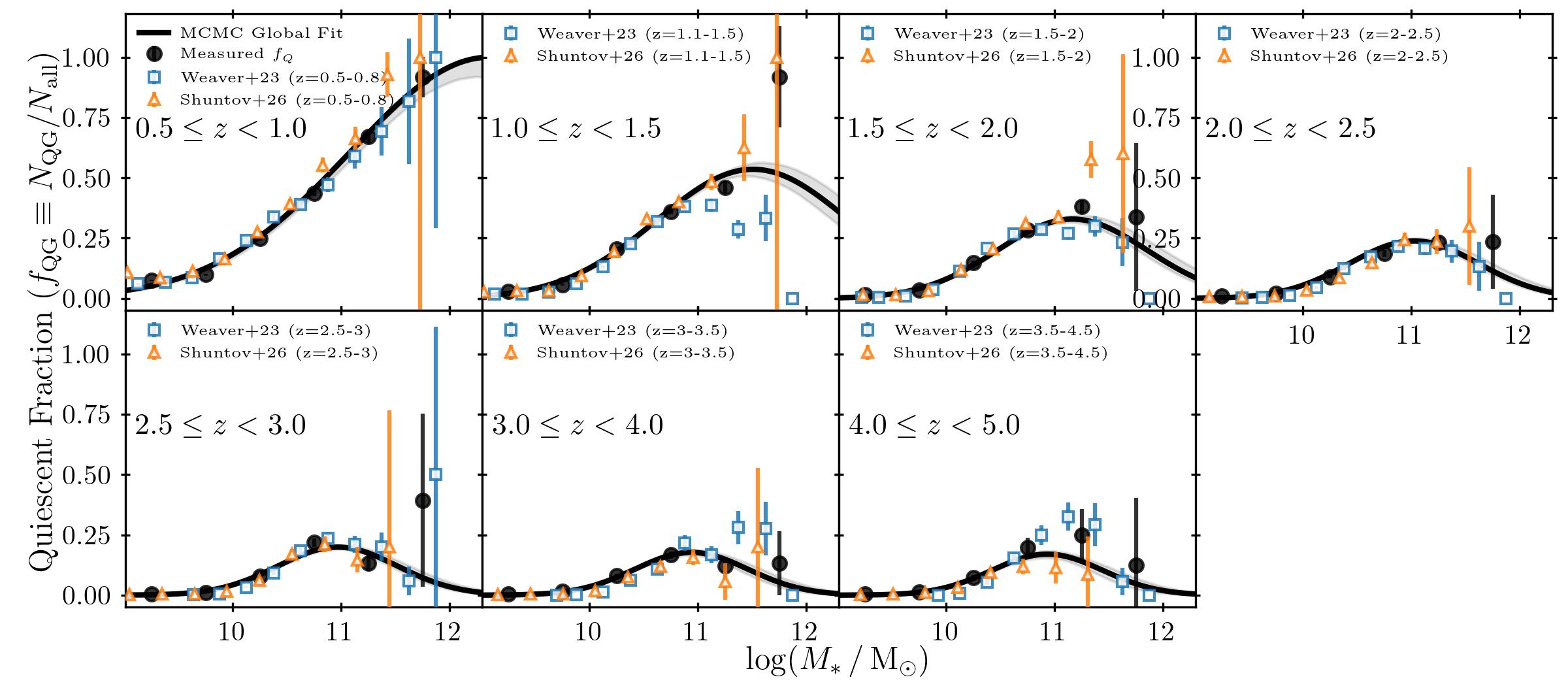}
\caption{Quiescent fraction as a function of stellar mass in the seven redshift intervals considered in this work. Black circles show our measured quiescent fractions, with the vertical error bars indicating their uncertainties. The solid black curves show the median global MCMC fit defined by Eqs.~\ref{eq:fq_model} and \ref{eq:fq_evolution}, while the gray shaded regions show the corresponding 16th--84th percentile posterior intervals. The global fit uses only measurements at ${\rm log}(M_\ast/{\rm M_\odot})\geq9.0$. For comparison, blue squares show the measurements of \citet{Weaver_2023}, and orange triangles show those of \citet{Shuntov_2025_GSMF}; the corresponding literature redshift intervals are given in the individual panel legends. The literature measurements are shown for comparison only and were not included in the fit.}
\label{fig:fq_global}
\end{figure*}

The aim of this part of the analysis was not to perform a separate, comprehensive study of the quiescent galaxies population, but to obtain a consistent estimate of the quiescent fraction required by the continuity-equation calculation presented in Sect.~\ref{sec:consist}. We therefore derived $f_{\rm Q}(x,z)$ from the same COSMOS-Web parent catalog and using the same quality selection, contaminant rejection, and $NUV-r-J$ classification adopted for the main-sequence analysis. This ensures that the main-sequence and quiescent-fraction terms describe the same galaxy populations.

We calculated the quiescent fractions in the redshift intervals $0.5\leq z<1.0$, $1.0\leq z<1.5$, $1.5\leq z<2.0$, $2.0\leq z<2.5$, $2.5\leq z<3.0$, $3.0\leq z<4.0$, and $4.0\leq z<5.0$. Within each redshift interval, the sample was divided into stellar-mass bins of width 0.5 dex over $7.5\leq x<12.0$. Galaxies were classified as quiescent using the $NUV-r-J$ criteria defined in Eq.~\ref{eq:nuvrj_selection}, without imposing an additional sSFR threshold. For each populated redshift--stellar-mass bin, the quiescent fraction was calculated as

\begin{equation}
f_{\rm Q}
=
\frac{N_{\rm Q}}{N_{\rm all}}
=
\frac{N_{\rm Q}}{N_{\rm Q}+N_{\rm SF}},
\label{eq:fq_definition}
\end{equation}

\noindent where $N_{\rm Q}$ and $N_{\rm SF}$ are the numbers of quiescent and star-forming galaxies, respectively (black dots in Fig.~\ref{fig:fq_global}).

We propagated the catalog redshift and stellar-mass uncertainties using 1000 Monte Carlo realizations. In each realization, the redshifts and stellar masses of the galaxies were perturbed using split Gaussian distributions whose lower and upper widths were set by the catalog-provided 16th and 84th percentiles. The galaxies were then reassigned to the redshift and stellar-mass bins and $f_{\rm Q}$ was recalculated. The dispersion of the resulting distribution in each bin was adopted as the corresponding $1\sigma$ uncertainty. The complete binned quiescent fractions, source counts, and uncertainties are available in machine-readable form in the associated online material.

\subsection{Mass and redshift dependence}
\label{sec:qg_func}

To obtain a continuous description of the quiescent fraction for use in the stellar-mass-function evolution model, we fitted all measurements at $x\geq9.0$ simultaneously. Using the logarithmic stellar-mass coordinate $x\equiv\log_{10}(M_\ast/{\rm M_\odot})$ defined in Sect.~\ref{sec:ms_func}, we adopted a Gaussian dependence on $x$,

\begin{equation}
f_{\rm Q}(x,z)
=
A(z)
\exp
\left[
-\frac{1}{2}
\left(
\frac{x-\mu(z)}
{\sigma(z)}
\right)^2
\right],
\label{eq:fq_model}
\end{equation}

\noindent where $A(z)$ determines the maximum quiescent fraction, $\mu(z)$ gives the logarithmic stellar mass at which this maximum occurs, and $\sigma(z)$ describes the width of the distribution. The redshift evolution of each of these three quantities was parametrized as

\begin{equation}
p(z)=a_p\exp(b_p z)+c_p,
\qquad
p\in\{A,\mu,\sigma\}.
\label{eq:fq_evolution}
\end{equation}

\noindent We fitted the resulting nine parameters simultaneously using the same Gaussian-likelihood MCMC approach described in Sect.~\ref{sec:ms_func}, adopting flat, bounded priors and sampling the posterior with \texttt{emcee}. The posterior medians and corresponding 16th--84th percentile intervals are listed in Table~\ref{tab:fq_fit}, while Fig.~\ref{fig:fq_global} shows the resulting global relation and its posterior uncertainty. The MCMC corner plot for the nine fitted parameters and the redshift evolution of $A$, $\mu$, and $\sigma$, are provided in the associated online material. The flattened posterior samples for all nine parameters are available in machine-readable form through the project GitHub repository.

\begin{table}
\begin{footnotesize}
\begin{center}
\caption{Best-fit parameters of the quiescent-fraction model of Eq.~\ref{eq:fq_model}. For each $p\in\{A,\mu,\sigma\}$, the redshift dependence is described by $p(z)=a_p\exp(b_pz)+c_p$. The quoted values are the posterior medians, with uncertainties corresponding to the 16th--84th percentile interval.}
\label{tab:fq_fit}
\setlength{\tabcolsep}{2.4mm}
\begin{tabular}{lccc}
\hline\hline
$p$ & $a_p$ & $b_p$ & $c_p$ \\
\hline
$A$ &
$\phantom{-}2.90_{-0.50}^{+0.55}$ &
$-1.65_{-0.16}^{+0.15}$ &
$\phantom{0}0.168_{-0.015}^{+0.014}$ \\
$\mu$ &
$\phantom{-}5.73_{-1.20}^{+1.38}$ &
$-1.85_{-0.31}^{+0.35}$ &
$10.937_{-0.076}^{+0.072}$ \\
$\sigma$ &
$\phantom{-}2.27_{-0.38}^{+0.45}$ &
$-1.44_{-0.29}^{+0.28}$ &
$\phantom{0}0.512_{-0.060}^{+0.055}$ \\
\hline
\end{tabular}
\end{center}
\end{footnotesize}
\end{table}

The fitted relation reproduces the strong evolution of both the normalization and the mass dependence of the quiescent fraction. At $z\lesssim1$, $f_{\rm Q}$ increases steeply across the observed stellar-mass range and approaches unity at the highest masses. Toward higher redshift, the maximum quiescent fraction decreases rapidly, while the mass corresponding to the maximum moves toward $x\sim11$. At $z\gtrsim3$, the fitted evolution approaches a relatively narrow, low-amplitude distribution centered close to this mass.

Fig.~\ref{fig:fq_global} also compares our measurements with the COSMOS2020 results of \citet{Weaver_2023} and the recent COSMOS-Web measurements of \citet{Shuntov_2026}. The overall mass and redshift trends are consistent with both studies. In particular, all measurements show a rapid increase in the quiescent fraction toward high stellar masses at low redshift and a progressively smaller quiescent contribution at earlier epochs.

Both \citet{Shuntov_2026} results and the analysis presented in this work are based on the COSMOS2025 catalog and use the same $NUV-r-J$ quiescent-galaxy selection. Nevertheless, small differences between the inferred $f_Q$ values can clearly be seen in Fig.\,\ref{fig:fq_global}. This can be attributed to a subtle methodological differences adopted, including slightly different redshift intervals. In addition, although the same parent catalog is used, the two analyses do not employ identical final galaxy selections. The stellar-mass-function analysis of \citet{Shuntov_2026} applies its own magnitude limit and the visual inspection of high-redshift galaxies, whereas our $f_{\rm Q}$ measurements include the starburst galaxies removal adopted for the present analysis. The modest point-to-point offsets seen in Fig.~\ref{fig:fq_global} are therefore consistent with the expected effects of the different binning and sample definitions and do not indicate a discrepancy between the two COSMOS-Web measurements.

\section{Stellar-mass-function consistency test}
\label{sec:consist}

\subsection{Motivation for the GSMF consistency test}
\label{sec:consist_why}

The main sequence determined in Sect.~\ref{sec:ms} describes the in-situ stellar-mass growth of star-forming galaxies, but by itself is not sufficient to describe the evolution of the total galaxy population, which is also affected by quiescent-galaxy fraction, stellar-mass return to the interstellar medium (ISM), and mergers. Because star formation increases $M_\ast$, the normalization and mass dependence of the main sequence determine how rapidly star-forming galaxies move through stellar-mass space, thereby strongly influencing resulting GSMF shape. The evolution of the mass function, therefore, provides a complementary constraint on the main-sequence relation: relatively small differences in the instantaneous SFR at fixed mass can accumulate over several Gyr and produce substantially different distributions.

We tested this consistency by initializing the total GSMF at $z\simeq 8$ and evolving it forward in time within a continuity-equation framework. The measured main sequence drives the in-situ growth of the star-forming population, while the mass- and redshift-dependent quiescent fraction derived in Sect.~\ref{sec:qg} accounts for galaxies that no longer undergo main-sequence growth. Material returned to the ISM through stellar evolution reduces the fraction of newly formed mass retained by galaxies, while mergers provide an additional way for galaxies to grow and require the corresponding removal of donor sources. The complete continuity-equation formulation and numerical implementation are described in Appendix~\ref{sec:consist_pipe}.

\subsection{GSMF evolution under different main-sequence prescriptions}
\label{sec:consist_test}

To test the sensitivity of the GSMF evolution to the adopted main sequence, we repeated the complete continuity calculation of Appendix~\ref{sec:consist_pipe} using several alternative main-sequence prescriptions. In every case, the same initial GSMF, stellar-mass return fraction, quiescent-fraction prescription, merger treatment, timestep, and numerical implementation were adopted. The main-sequence relation was therefore the only component varied between the calculations, allowing differences in the resulting GSMFs to be traced directly to differences in the assumed SFR--$M_\ast$ function.

For the relation derived in this work, uncertainties in the evolved GSMF were propagated using Monte Carlo realizations of the complete continuity calculation. In each realization, we drew one complete parameter set from the main-sequence posterior chain and one from the $f_{\rm Q}$ posterior chain, together with the initial Schechter GSMF of \citet{Shuntov_2025_GSMF} drawn using its asymmetric parameter uncertainties. The median evolved GSMF and its 16th--84th percentile uncertainty were then calculated at each stellar mass and redshift. The alternative main-sequence prescriptions were evolved deterministically and are therefore represented by individual realizations without uncertainty ranges. The resulting GSMF evolution is presented and compared with the observed COSMOS-Web GSMFs in Sect.~\ref{sec:disc_mass_grow}.

\section{Discussion}
\label{sec:disc}

\begin{figure*}
\centering
\includegraphics[width=\textwidth]{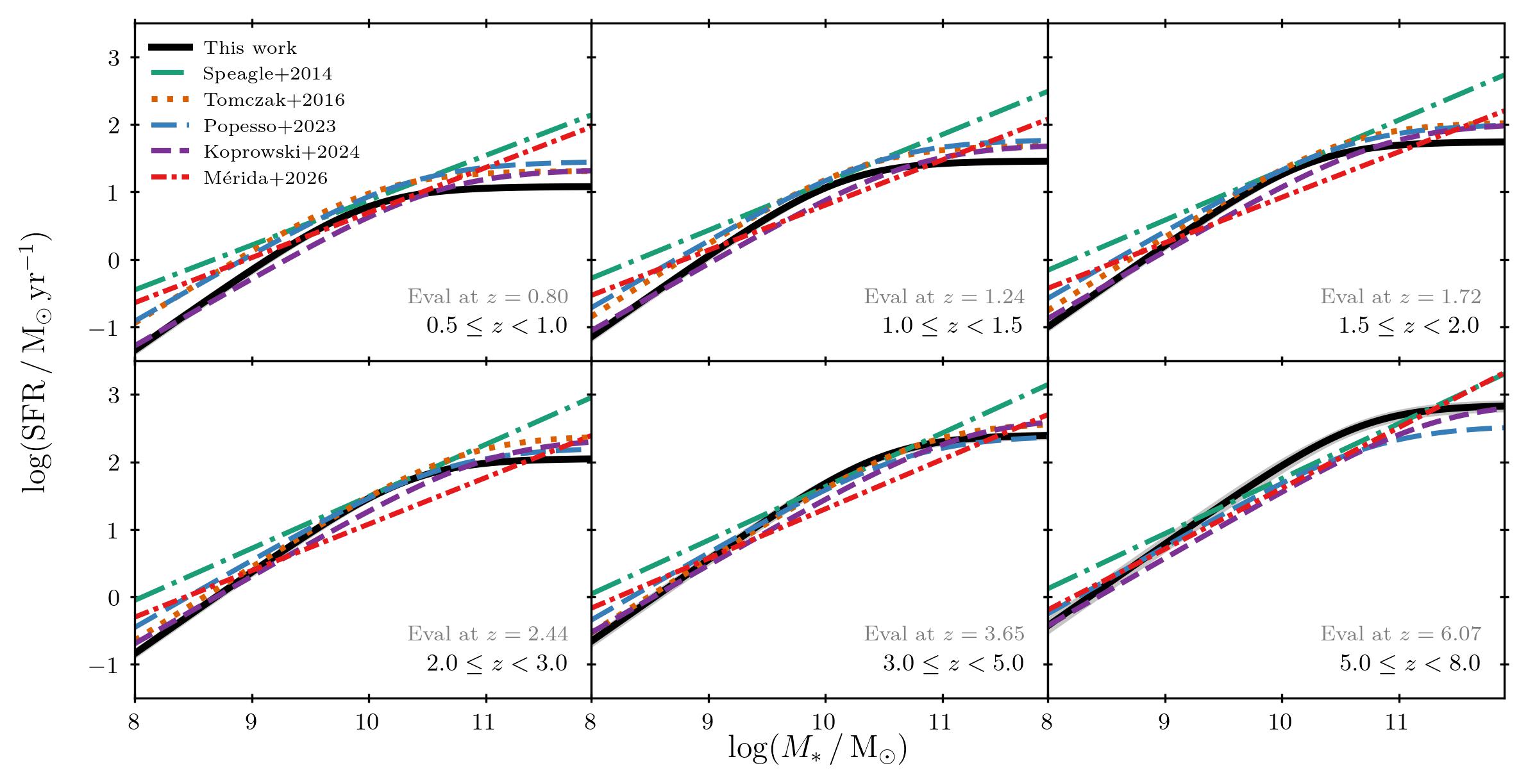}
\caption{Comparison of the star-forming main-sequence relation derived in this work with the prescriptions of \citet{Speagle_2014}, \citet{Tomczak_2016}, \citet{Popesso_2023}, \citet{Koprowski_2024}, and \citet{Merida_2026}. The relations are evaluated at the representative redshift indicated in each panel, corresponding to the six redshift intervals adopted for our measurements. The solid black curves show the relation derived in this work, with the gray shaded regions indicating its posterior uncertainty. For \citet{Merida_2026}, the redshift evolution used to evaluate the relation between and below their measured redshift intervals is described in Appendix~\ref{sec:merida_ms_evolution}.}
\label{fig:ms_models_comp}
\end{figure*}

\subsection{A mass-complete JWST main sequence at high redshift}
\label{sec:disc_ms}

\subsubsection{Shape of the MS}
\label{sec:disc_ms_shape}

As shown in Fig.~\ref{fig:ms_fit}, the star-forming main sequence retains a similar overall shape across the full redshift range considered here: the SFR increases approximately as a power law with stellar mass below a characteristic turnover mass and progressively flattens above it, while the normalization increases strongly toward higher redshift. The turnover mass also increases with redshift, so that the flattening occurs at progressively larger stellar masses at earlier epochs. Importantly, because quiescent galaxies were removed before stacking, the high-mass flattening cannot simply be attributed to the increasing contribution of the quenched population.

The low-mass slope is well constrained by the depth of the JWST-selected sample, which extends the mass-complete measurements considerably below the turnover. We obtain $\gamma=1.215\pm0.028$, corresponding to $\mathrm{SFR}\propto M_\ast^{1.215}$ in the low-mass regime. Consequently, the sSFR is not strictly independent of stellar mass but scales approximately as $\mathrm{sSFR}\propto M_\ast^{0.215}$, implying slightly faster fractional stellar-mass growth toward higher masses below the turnover. This differs from the nearly linear low-mass relation, $\gamma\simeq1$, frequently assumed in previous studies.

The general form of the relation can be understood in terms of the gas supply available to main-sequence galaxies. Previous studies showed that main-sequence galaxies approximately follow an integrated Schmidt--Kennicutt relation, while their molecular-gas fractions vary systematically with stellar mass and redshift \citep{Santini_2014,Wang_2022}. The increasing normalization of the MS toward high redshift is therefore consistent with the increasing gas fractions of galaxies at earlier epochs. At the high-mass end, the flattening indicates that the available gas supply or its conversion into stars does not increase proportionally with stellar mass; suppression of cold-gas accretion above sufficiently massive halos has been proposed as one possible origin of this behavior \citep{Daddi_2022}, although changes in star-formation efficiency may also contribute \citep{Schreiber_2016}.

\subsubsection{Comparison to recent literature}
\label{sec:disc_ms_comp}

Fig.~\ref{fig:ms_models_comp} compares our best-fitting main sequence with the relations of \citet{Speagle_2014}, \citet{Tomczak_2016}, \citet{Popesso_2023}, \citet{Koprowski_2024}, and \citet{Merida_2026}, evaluated at representative redshifts within each of our six bins. Although the different prescriptions broadly agree with each other, some differences are apparent in their normalization, low-mass slope, turnover mass, and degree of high-mass flattening. A particularly important inconsistency can be seen at low stellar masses, where our data show $\gamma=1.215\pm0.028$, whereas most of the comparison relations are approximately linear or sub-linear. In particular, both \citet{Speagle_2014} and \citet{Merida_2026} find an intrinsic slope significantly below 1, substantially shallower than our relation. The consequences of these different slopes for the long-term evolution of the low-mass GSMF are discussed in Sect.~\ref{sec:disc_mass_grow}.

Both our sample and the data used in \citet{Merida_2026} are based on the JWST observations extending to low stellar masses. The inconsistencies between both MS shapes can be attributed to the methodological differences adopted in both works. \citet{Merida_2026} determine SFRs from \texttt{Dense Basis} SED fits with non-parametric star-formation histories, whereas our values combine the unobscured UV emission with the obscured component measured directly from stacked FIR and submillimeter data. In addition, their SFR-completeness limits at a given redshift are derived from the rest-frame UV detection limits using either $A_V=0$ or a single median value. Because dust attenuation varies systematically with stellar mass \citep[e.g.][]{McLure_2018,Koprowski_2018,Wijesekera_2026}, such a redshift-only correction may introduce a mass-dependent selection effect: in particular, if the adopted median attenuation exceeds the typical value of low-mass galaxies, the corresponding intrinsic SFR limit will be overestimated and a fraction of lower-SFR low-mass systems may be excluded from the final sample, potentially flattening the inferred MS relation.

Some inconsistencies at higher stellar masses, on the other hand, can be attributed to the different functional forms assumed. The \citet{Speagle_2014} relation is a single power law and therefore has no high-mass turnover. Similarly, the all-galaxy \citet{Merida_2026} relation adopted here was fitted as a single power law and consequently cannot reproduce a high-mass flattening by construction. The remaining prescriptions show varying degrees of flattening, with modest offsets that can arise from differences in SFR indicators, sample selection, and the treatment of dust emission. In particular, FIR-based SFR estimates remain sensitive to assumptions about the dust temperature when the SED is not directly constrained on both sides of its peak \citep[e.g.][]{Koprowski_2024}. The interpolation and extrapolation used to construct a continuous redshift-dependent representation of the \citet{Merida_2026} relation are described in Appendix~\ref{sec:merida_ms_evolution}.

\subsection{Implications for stellar-mass growth}
\label{sec:disc_mass_grow}

\begin{figure*}
\centering
\includegraphics[width=\textwidth]{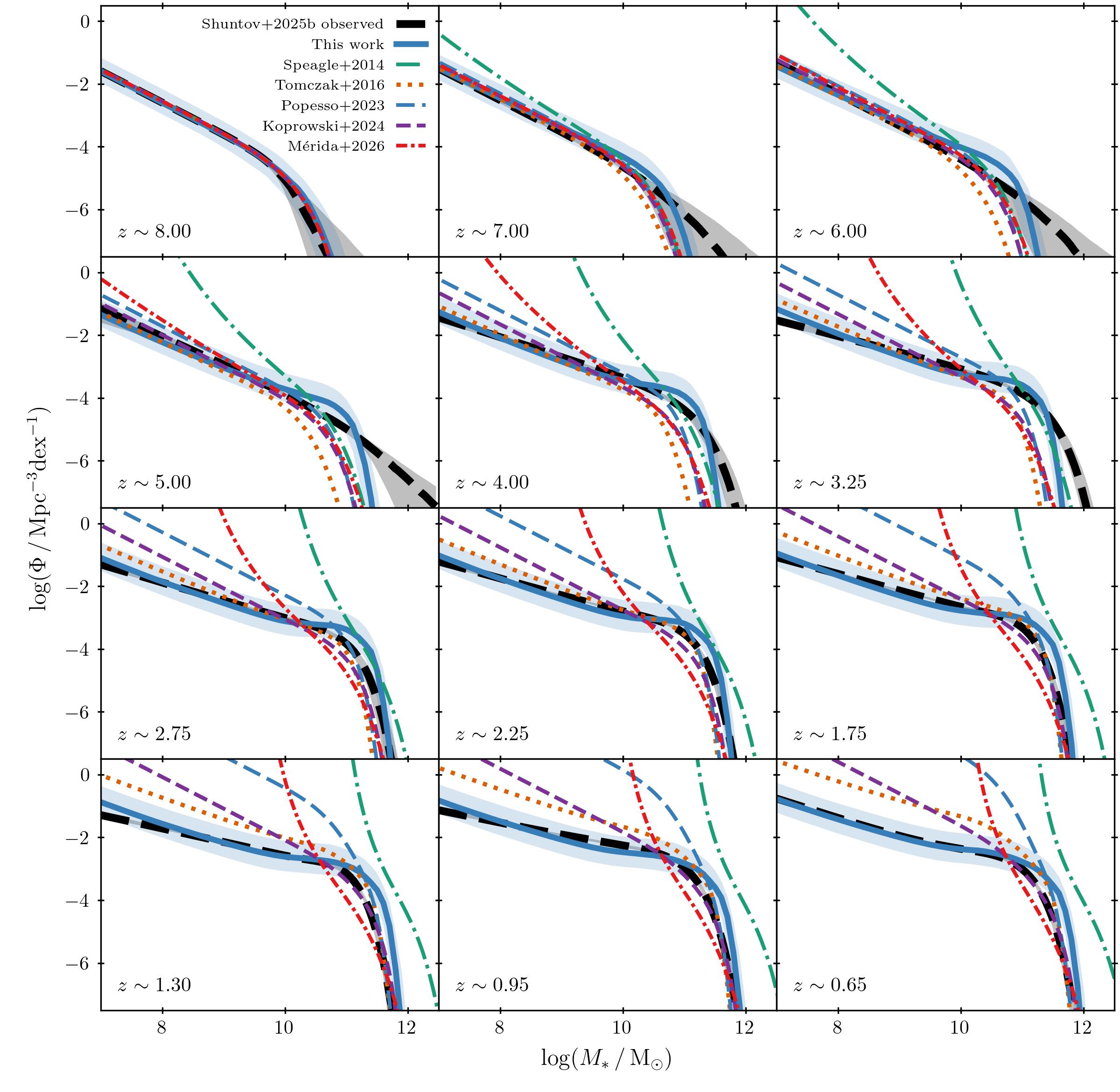}
\caption{Evolution of the total GSMF for different adopted star-forming main-sequence prescriptions. The black dashed curves and gray shaded regions show the observed COSMOS-Web GSMFs and their 16th--84th percentile ranges from \citet{Shuntov_2025_GSMF}. The blue curves show the median GSMFs evolved using the main-sequence relation derived in this work, with the blue shaded regions indicating the 16th--84th percentile ranges (Sect.~\ref{sec:consist_test}). The remaining curves show the deterministic results obtained using the main-sequence prescriptions of \citet{Speagle_2014}, \citet{Tomczak_2016}, \citet{Popesso_2023}, \citet{Koprowski_2024}, and \citet{Merida_2026}. All models were evolved from the same initial GSMF using otherwise identical assumptions. The differences between the adopted main-sequence prescriptions and the resulting GSMF evolution are discussed in Sect.~\ref{sec:disc_mass_grow}.}
\label{fig:gsmf_evol_ms}
\end{figure*}

To examine how the adopted main sequence affects stellar-mass growth, we evolved each of the relations shown in Fig.~\ref{fig:ms_models_comp}, from the same initial $z\simeq8$ GSMF using the continuity framework described in Appendix~\ref{sec:consist_pipe}. The result of this exercise is depicted in Fig.~\ref{fig:gsmf_evol_ms}. Because the initial GSMF, quiescent fraction, stellar-mass return, merger prescription, and numerical treatment were kept fixed, differences between the evolved GSMFs can be attributed directly to the adopted main-sequence prescription.

The normalization of the main sequence determines the overall rate at which galaxies move toward higher stellar masses, while its mass dependence determines how this growth varies across the galaxy population. The turnover mass and degree of high-mass flattening regulate the in-situ buildup of the high-mass end, and their redshift evolution determines when these effects become most significant. The low-mass evolution is also sensitive to the MS slope. Below the turnover, our relation follows approximately $\mathrm{SFR}\propto M_\ast^\gamma$, with $\gamma=1.215\pm0.028$. Since the velocity through logarithmic stellar-mass space scales approximately as $v_x\propto M_\ast^{\gamma-1}$, this produces differential growth across the low-mass population and allows the initially steep GSMF to flatten progressively toward lower redshift. The resulting evolution closely follows that seen in the observed COSMOS-Web GSMFs. Mergers have only a modest influence in this regime, changing the low-mass number densities by $\lesssim0.2$ dex by $z\simeq0.65$ without substantially affecting the slope.

The comparison with the literature further illustrates how modest differences in the MS propagate into substantially different GSMF evolution. The \citet{Speagle_2014} relation is a single power law with a slope that remains below unity, so its differential mass growth acts in the opposite sense to our $\gamma>1$ relation and drives the low-mass GSMF far above the observed one by $z\simeq0.65$. A similar behavior is seen for \citet{Merida_2026}, whose all-galaxy fits have sub-linear slopes of $\beta\simeq0.7$--$0.8$ over most of the redshift range. The resulting GSMF increasingly exceeds the observed low-mass number densities toward lower redshift and fails to reproduce their observed flattening. The \citet{Popesso_2023} relation is much closer to ours in shape, but its slightly higher normalization produces systematically faster growth and an evolved GSMF that also lies well above the observations by low redshift, showing how small normalization offsets accumulate over several Gyr. The \citet{Tomczak_2016} relation has $\gamma=1.091$, but its quadratic redshift parameterization causes both the characteristic mass and normalization to turn over and decline when extrapolated beyond $z\sim4$. As a result, extending the prescription to our $z\simeq8$ starting point produces an artificially low main-sequence normalization and the non-monotonic behavior seen in the evolved GSMFs. Because of the limited stellar-mass dynamic range, \citet{Koprowski_2024} fixed the low-mass MS slope to $\gamma=1$, which results in the shape of the evolved GSMF in this regime to remain approximately unchanged throughout the simulation.

At high stellar masses, the interpretation is less straightforward. The turnover and subsequent flattening of the main sequence reduce the relative importance of in-situ star formation, while merger-driven growth becomes increasingly important. Although the GSMF evolved using our relation remains broadly consistent with the observations, some discrepancies are visible around the high-mass end. Potential explanations include the extrapolation of the \citet{Rodriguez_2016} merger prescription beyond its calibrated redshift range at $z>4$. Residual starburst contamination may also contribute, since A3COSMOS is constructed from inhomogeneous archival ALMA observations obtained by individual PI-led programs and has an undetermined selection function \citep{Liu_2019}. For this reason, A3COSMOS data cannot be treated as a luminosity-complete census of starbursts, with a fraction of such systems likely remaining in our stacking sample.

Overall, Fig.~\ref{fig:gsmf_evol_ms} shows how small differences in the shape of different MS functions can drive significant inconsistencies in the long-term stellar-mass evolution. The effects of slightly different low-mass slopes and normalizations become strongly amplified when integrated over several Gyr. The observed evolution of the GSMF, therefore, becomes a sensitive and largely independent test of the mass and redshift dependence of the star-forming main sequence.

\section{Summary}
\label{sec:summ}

We have used the COSMOS-Web catalog to determine the mass-complete star-forming main sequence out to $z=8$ and to test whether the resulting stellar-mass growth is consistent with the observed evolution of the GSMF. We stacked the mass-selected star-forming galaxies in the FIR \textit{Herschel} and JCMT maps, correcting the recovered flux densities for blending using forward-modeled maps, and combined the resulting IR luminosities with the unobscured UV emission to derive total SFRs. We fitted the resulting SFR--$M_\ast$ measurements with a redshift-dependent MS relation and independently measured the quiescent fraction from the same COSMOS-Web population. Finally, we initialized the observed COSMOS-Web GSMF at $z=\simeq8$ and evolved it forward within a continuity-equation framework including in-situ stellar-mass growth, stellar-mass return, the quiescent fraction, and merger-driven mass growth and donor destruction. The main findings of this work can be summarized as follows:

\begin{enumerate}[label=(\roman*)]

\item The star-forming MS is well described by a power law at low stellar masses that progressively flattens above a characteristic turnover mass. Its normalization increases strongly toward higher redshift, while the turnover shifts toward higher stellar masses. The substantially greater stellar-mass dynamic range provided by COSMOS-Web allows the low-mass slope to be constrained directly, giving $\gamma=1.215\pm0.028$. The corresponding $\mathrm{sSFR}\propto M_\ast^{0.215}$ dependence implies that galaxies at progressively larger masses below the turnover undergo slightly faster fractional stellar-mass growth.

\item The quiescent fraction shows a strong dependence on both stellar mass and redshift. At low redshift, the quiescent contribution rises rapidly toward high stellar masses, while its overall amplitude decreases toward earlier epochs. We fitted these measurements with a continuous $f_{\rm Q}(M_\ast,z)$ relation and used it to reduce the population-averaged in-situ growth of the total GSMF. This provides a consistent treatment of the relative star-forming and quiescent contributions without fitting the MS to the observed GSMF evolution.

\item Evolving the COSMOS-Web GSMF from $z\simeq8$ using the MS measured in this work reproduces the observed evolution substantially better than the other MS prescriptions considered, particularly at low stellar masses. The low-mass behavior follows directly from the measured $\gamma>1$: the resulting mass-dependent fractional growth progressively changes the shape of the GSMF toward lower redshift, consistent with the observed flattening. Mergers have only a modest effect in this regime, changing the low-mass number densities by $\lesssim0.2$ dex by $z\simeq0.65$ without substantially affecting their slope. The evolution of the low-mass GSMF is therefore driven primarily by the mass dependence of the MS.

\item The comparison with recent literature MS functions demonstrates that relatively small variations in their shapes can lead to large differences after several Gyr of stellar-mass growth. The sub-linear slopes of \citet{Speagle_2014} and \citet{Merida_2026} produce a strong excess of low-mass galaxies, failing to reproduce the observed flattening of the observed GSMF by $z\simeq0.65$. The slightly higher normalization of \citet{Popesso_2023} leads to systematically faster growth and an evolved mass function lying substantially above the observations at low redshift. The \citet{Tomczak_2016} relation requires extrapolation of its quadratic redshift dependence beyond its calibrated range of $z\sim4$, resulting in an artificially low normalization near our $z\simeq8$ starting point. The \citet{Koprowski_2024} relation produces an evolution closer to ours, although fixing its low-mass slope to $\gamma=1$ keeps the shape of the evolved mass function in this regime approximately constant. These results show that the observed GSMF provides a sensitive independent test of the shape and normalization of the star-forming MS.

\end{enumerate}

\section*{Data Availability}
\label{sec:data_avail}

The machine-readable tables associated with this work are available in the \href{https://github.com/maciejpkoprowski/COSMOS-Web-GSMF-MS-data}{project GitHub repository}. The supplementary figures and diagnostic plots are available in the associated \href{https://doi.org/10.5281/zenodo.21648421}{Zenodo record}.

\begin{acknowledgements}
This research was funded in whole or in part by the National Science Centre, Poland (grant no. 2023/50/E/ST9/00383). For the purpose of Open Access, the author has applied a CC-BY public copyright license to any Author Accepted Manuscript (AAM) version arising from this submission. K.L. acknowledges the support of the National Science Centre, Poland, through the PRELUDIUM grant UMO-2023/49/N/ST9/00746.
\end{acknowledgements}

\bibliographystyle{aa}
\bibliography{my_papers}

\begin{appendix}

\section{Continuity-equation framework}
\label{sec:consist_pipe}

We described the evolution of the total GSMF in terms of the logarithmic stellar-mass coordinate $x\equiv\log_{10}(M_\ast/{\rm M_\odot})$ and the differential number density $\Phi(x,t)\equiv {\rm d}N/({\rm d}V\,{\rm d}x)$. In continuous form, the evolution corresponding to our calculation can be written as

\begin{equation}
\frac{\partial \Phi(x,t)}{\partial t}
+
\frac{\partial}{\partial x}
\left[
\Phi(x,t)\,v_x(x,t)
\right]
=
-\mathcal{D}_{\rm merg}[\Phi](x,t),
\label{eq:gsmf_continuity}
\end{equation}

\noindent where $\mathcal{D}_{\rm merg}[\Phi]$ denotes the reduction in number density caused by the destruction of lower-mass donor galaxies in mergers, and the velocity through logarithmic stellar-mass space is

\begin{equation}
v_x(x,t)
=
\frac{\dot{M}_{\ast,\rm insitu}(x,z)
+\dot{M}_{\ast,\rm merg}(x,z)}
{M_\ast(x)\ln 10},
\label{eq:gsmf_mass_velocity}
\end{equation}

\noindent where $M_\ast(x)=10^x\,{\rm M_\odot}$ and $z=z(t)$ is determined from the adopted cosmology. The first term in the numerator describes stellar-mass growth through star formation, after accounting for stellar mass return and for the fraction of galaxies that are quiescent, while the second describes stellar mass accreted through mergers. These components are described separately in Sects.~\ref{sec:consist_pipe_ms}--\ref{sec:consist_pipe_merg}.

We divided the GSMF into logarithmic stellar-mass bins, each containing an integrated number density $N_i$, and made the bin boundaries move as galaxies increased their stellar mass. Here, $i$ denotes a mass bin, $M_{i-1/2}$ and $M_{i+1/2}$ are its lower and upper boundaries, respectively, and $n$ denotes the timestep. At the start of the calculation, $N_i=\Phi(x_i)\Delta x_i$. During each timestep, the bin boundaries were shifted according to the combined in-situ and merger-driven stellar-mass increments, while the integrated number densities were changed only by merger donor destruction. This can be summarized as

\begin{equation}
M_{i\pm 1/2}^{\,n+1}
=
M_{i\pm 1/2}^{\,n}
+
\Delta M_{\ast,\rm insitu,\,i\pm 1/2}^{\,n}
+
\Delta M_{\ast,\rm merg,\,i\pm 1/2}^{\,n},
\label{eq:gsmf_edges_update}
\end{equation}

\begin{equation}
N_i^{\,n+1}
=
N_i^{\,n}
-
\Delta N_{{\rm donor},i}^{\,n}.
\label{eq:gsmf_number_update}
\end{equation}

\noindent where the stellar-mass increments in Eq.~\ref{eq:gsmf_edges_update} are evaluated at the corresponding moving bin boundaries. The donor-removal procedure explicitly limits the number of destroyed galaxies to the number available in each bin, ensuring $\Delta N_{{\rm donor},i}\leq N_i$. The GSMF was reconstructed after each timestep as $\Phi_i=N_i/\Delta x_i$, where $\Delta x_i=x_{i+1/2}-x_{i-1/2}$.

The evolution was carried out forward in cosmic time from $z=8$, using the analytical single-Schechter fit of \citet{Shuntov_2025_GSMF} in the $7.5\leq z<8.5$ interval as the initial GSMF. Cosmic time was calculated using the adopted cosmology and the evolution was performed with a timestep of 20\,Myr. The main-sequence, quiescent-fraction, and merger quantities were evaluated at the beginning of each timestep before the corresponding stellar-mass and donor-number updates were applied.

\subsection{In-situ stellar-mass growth along the main sequence}
\label{sec:consist_pipe_ms}

For each stellar mass and redshift, the main-sequence relation specifies the instantaneous star-formation rate of a star-forming galaxy. For the relation derived in Sect.~\ref{sec:ms_func}, this is obtained from $\psi(x,z)$ as

\begin{equation}
\mathrm{SFR}_{\rm MS}(x,z)
=
10^{\psi(x,z)}\,{\rm M_\odot}\,\mathrm{yr}^{-1}.
\label{eq:consist_sfr_ms}
\end{equation}

\noindent However, not all of the newly formed stellar mass remains locked into stars, because a fraction is returned to the interstellar medium through stellar evolution. Following \citet{Leja_2015}, we adopted a constant stellar-mass return fraction of $R=0.36$. The retained stellar-mass growth rate of a galaxy on the main sequence is therefore

\begin{equation}
\dot{M}_{\ast,\rm SF}(x,z)
=
(1-R)\,\mathrm{SFR}_{\rm MS}(x,z)
=
0.64\,\mathrm{SFR}_{\rm MS}(x,z).
\label{eq:consist_ms_growth}
\end{equation}

\noindent Although $\dot{M}_{\ast,\rm SF}$ is evaluated as a function of the logarithmic stellar-mass coordinate $x$, it is a linear mass-growth rate in ${\rm M_\odot}\,\mathrm{yr}^{-1}$, with $M_\ast(x)=10^x\,{\rm M_\odot}$. During each timestep $\Delta t$, the corresponding stellar-mass increment is $\Delta M_{\ast,\rm SF}=\dot{M}_{\ast,\rm SF}\Delta t$, evaluated at the moving mass-bin boundaries. The reduction of this growth caused by the presence of quiescent galaxies is described in Sect.~\ref{sec:consist_pipe_qg}.

\subsection{Quiescent-fraction suppression of in-situ mass growth}
\label{sec:consist_pipe_qg}

The growth rate in Eq.~\ref{eq:consist_ms_growth} applies only to galaxies that are forming stars on the main sequence, whereas the total GSMF contains both star-forming and quiescent galaxies. We therefore used the quiescent fraction derived in Sect.~\ref{sec:qg_func} to calculate the population-averaged in-situ stellar-mass growth rate as

\begin{equation}
\dot{M}_{\ast,\rm insitu}(x,z)
=
\left[1-f_{\rm Q}(x,z)\right]
\dot{M}_{\ast,\rm SF}(x,z).
\label{eq:consist_fq_growth}
\end{equation}

\noindent Thus, if all galaxies at a given mass and redshift are star forming, $f_{\rm Q}=0$ and the full main-sequence growth rate is applied, while increasing $f_{\rm Q}$ progressively reduces the average in-situ growth. The fitted $f_{\rm Q}(x,z)$ was restricted to the physical range $0\leq f_{\rm Q}\leq1$ and evaluated at the current redshift and moving mass-bin boundaries at each timestep. The corresponding mass increment was then $\Delta M_{\ast,\rm insitu}=\dot{M}_{\ast,\rm insitu}\Delta t$.

In this treatment, quiescent galaxies are not removed from the total GSMF and galaxies are not explicitly transferred between star-forming and quiescent populations. Instead, $f_{\rm Q}$ reduces the mean in-situ stellar-mass growth rate of the full population at each mass and redshift. The number density in a mass bin is therefore unchanged by this term; changes in galaxy number arise only from merger donor destruction described in Sect.~\ref{sec:consist_pipe_merg}.

\subsection{Merger/donor-destruction term}
\label{sec:consist_pipe_merg}

We accounted for stellar-mass growth through mergers using the specific stellar-mass accretion rate derived from the Illustris simulation by \citet{Rodriguez_2016}. This quantity gives the average stellar mass accreted through mergers per unit galaxy stellar mass, merger mass ratio, and time,

\begin{equation}
\dot{m}_{\rm acc,\ast}(M_\ast,\mu,z)
=
\frac{1}{M_\ast}
\frac{{\rm d}M_{\rm acc}}{{\rm d}\mu\,{\rm d}t},
\label{eq:merger_specific_rate}
\end{equation}

\noindent where $\mu=M_{\ast,\rm donor}/M_{\ast,\rm primary}$ is the stellar-mass ratio of the merger. We adopted the fitting function of \citet{Rodriguez_2016}:

\begin{equation}
\begin{split}
\dot{m}_{\rm acc,\ast}(M_\ast,\mu,z)
={}&
A(z)
\left(\frac{M_\ast}{10^{10}\,{\rm M_\odot}}\right)^{\alpha(z)}
\left[
1+
\left(\frac{M_\ast}{2\times10^{11}\,{\rm M_\odot}}\right)^{\delta(z)}
\right] \\
&\times
\mu^{\beta(z)+\gamma\log_{10}\left(M_\ast/10^{10}\,{\rm M_\odot}\right)}
\frac{\mu}{1+3\mu},
\end{split}
\label{eq:merger_kernel}
\end{equation}

\noindent with $A(z)=A_0(1+z)^\eta$, $\alpha(z)=\alpha_0(1+z)^{\alpha_1}$, $\beta(z)=\beta_0(1+z)^{\beta_1}$, and $\delta(z)=\delta_0(1+z)^{\delta_1}$, adopting the best-fitting parameters given in their paper. We included mergers over $0.01\leq\mu\leq1$, thereby accounting for major, minor, and very minor mergers while excluding mass ratios for which the contribution to stellar-mass growth is negligible. The corresponding merger-driven stellar-mass growth rate of Eq.~\ref{eq:gsmf_mass_velocity} was obtained by integrating over the adopted mass-ratio interval,

\begin{equation}
\dot{M}_{\ast,\rm merg}(M_\ast,z)
=
M_\ast
\int_{0.01}^{1}
\dot{m}_{\rm acc,\ast}(M_\ast,\mu,z)\,{\rm d}\mu.
\label{eq:merger_mass_growth}
\end{equation}

To determine which donor mass bins supplied this accreted mass, we evaluated the contribution of the same function over the mass-ratio interval corresponding to each donor bin. For example, a donor bin spanning $\mu_{\rm low,j} M_{\rm primary}<M_{\rm donor}<\mu_{\rm high,j} M_{\rm primary}$ contributes

\begin{equation}
\dot{M}_{\ast,\rm merg}^{\Delta\mu_j}
=
M_{\rm primary}
\int_{\mu_{\rm low,j}}^{\mu_{\rm high,j}}
\dot{m}_{\rm acc,\ast}(M_{\rm primary},\mu,z)\,{\rm d}\mu.
\end{equation}

\noindent The fraction of the total merger-driven growth supplied by that donor-mass range is therefore
\begin{equation}
f_{\Delta\mu_j}
=
\frac{
\int_{\mu_{\rm low,j}}^{\mu_{\rm high,j}}
\dot{m}_{\rm acc,\ast}(M_{\rm primary},\mu,z)\,{\rm d}\mu
}{
\int_{0.01}^{1}
\dot{m}_{\rm acc,\ast}(M_{\rm primary},\mu,z)\,{\rm d}\mu
}.
\end{equation}

\noindent The total stellar mass accreted by the primary bin during a timestep was then divided among all eligible donor bins according to these relative contributions. In the numerical implementation, the integral over each donor bin was approximated as $\dot{m}_{\rm acc,\ast}(M_{\rm primary},\mu_j,z)\,\Delta\mu_j$, where $\mu_j$ and $\Delta\mu_j$ are the representative mass ratio and width of that donor bin. The stellar mass assigned to each donor bin was then converted into a number of destroyed galaxies by dividing it by the characteristic stellar mass of that bin. If the requested donor number exceeded the number of galaxies available in a bin, the removal was limited to the available population and the associated receiver growth was reduced accordingly.

The mass of donor galaxies was added to the primary population through $\Delta M_{\ast,\rm merg}$ in Eq.~\ref{eq:gsmf_edges_update}, while the destroyed donor galaxies reduced $N_i$ through $\Delta N_{{\rm donor},i}$ in Eq.~\ref{eq:gsmf_number_update}. This construction therefore couples merger-driven stellar-mass growth to the corresponding decrease in galaxy number while conserving the stellar mass exchanged between donor and primary populations.

\section{Redshift parameterization of the Mérida et al. main sequence}
\label{sec:merida_ms_evolution}

For the comparison with \citet{Merida_2026}, we adopted their intrinsic all-galaxy fits obtained with asymmetric stellar-mass and SFR uncertainties. They fitted their main-sequence relation independently in eight redshift intervals over $1<z<9$ as $\log_{10}(\mathrm{SFR}/{\rm M_\odot}\,\mathrm{yr}^{-1})=\alpha+\beta x$, where $x\equiv\log_{10}(M_\ast/{\rm M_\odot})$, and no continuous redshift dependence was provided. To obtain a smooth relation for the MS comparison and continuity calculation, we first expressed each fit about the pivot $x_{\rm p}=9.5$ as $\log_{10}(\mathrm{SFR}/{\rm M_\odot}\,\mathrm{yr}^{-1})=A_{9.5}(z)+\beta(z)(x-9.5)$, where $A_{9.5}=\alpha+9.5\beta$. This avoids fitting the strongly slope-dependent intercept $\alpha$, which is defined at $x=0$ far outside the observed stellar-mass range.

We fitted the eight tabulated central values of $A_{9.5}$ and $\beta$ independently with second-order polynomials in redshift, obtaining $A_{9.5}(z)=-0.02079z^2+0.29422z+0.13840$ and $\beta(z)=0.00858z^2-0.01461z+0.67417$. These functions were used only to provide a smooth representation of the published discrete relations and were not refitted to the galaxy measurements. The relation was extrapolated below the \citet{Merida_2026} range of $z>1$ where required by our comparison, down to the lowest redshift used in the GSMF evolution, $z\simeq0.65$. No corresponding high-redshift extrapolation is required because the GSMF evolution begins at $z=8$. A diagnostic showing the original parameter values together with the adopted redshift dependence and its low-redshift extrapolation is provided in the associated online material.

\end{appendix}

\end{document}